\documentclass[11pt]{article}

\usepackage[final]{acl}

\usepackage{times}
\usepackage{latexsym}

\usepackage[T1]{fontenc}
\usepackage[utf8]{inputenc}

\usepackage{microtype}

\usepackage{inconsolata}

\usepackage{graphicx}
\usepackage{amsmath,amssymb,amsfonts}
\usepackage{multirow}
\usepackage{stfloats} % improves placement of full-width table*/figure* floats
\usepackage{pgfplots}
\pgfplotsset{compat=1.18}
\definecolor{seriesone}{HTML}{2A78D6}
\definecolor{seriestwo}{HTML}{EB6834}
\definecolor{seriesthree}{HTML}{1BAF7A}
\definecolor{seriesgrid}{HTML}{D5D5D2}
\definecolor{seriesaxis}{HTML}{8A8A85}

\newcommand{\emailfont}{\fontsize{10}{12}\selectfont}

\title{Disentangled Global-Local Feature Learning with E-Branchformer for Audio Deepfake Detection}

\author{
  Phuong Tuan Dat\textsuperscript{1,}\thanks{\ \ Equal contribution.} \quad
  Ho Bao Thu\textsuperscript{2,}\footnotemark[1] \quad
  Nguyen Tran Trung\textsuperscript{2} \quad
  Pham Viet Hoang\textsuperscript{2} \quad
  Nguyen Thi Thu Trang\textsuperscript{2} \\[2pt]
  \normalfont
  \textsuperscript{1}Department of Electrical and Computer Engineering,
  National University of Singapore \\
  \textsuperscript{2}School of Communication and Information Technology,
  Hanoi University of Science and Technology \\[3pt]
  \emailfont
  \texttt{phuongtuandat@u.nus.edu} \quad
  \texttt{thu.hb226003@sis.hust.edu.vn} \quad
  \texttt{trung.nt2400117@sis.hust.edu.vn} \\
  \emailfont
  \texttt{hoang.pv224854@sis.hust.edu.vn} \quad
  \texttt{trangntt@soict.hust.edu.vn}
}

\begin{document}
\maketitle

\begin{abstract}
The rapid advancement of voice synthesis technologies such as text-to-speech and voice conversion poses significant threats to speech-based authentication systems, necessitating robust deepfake detection methods. In this work, we propose a novel E-Branchformer-based architecture that effectively leverages self-supervised speech representations for audio deepfake detection. Our model employs parallel branches to simultaneously capture global contextual dependencies through multi-head self-attention and local temporal patterns through convolutional processing. To enhance discriminative capability, we integrate depthwise convolution and Squeeze-and-Excitation modules that enrich the classification token with refined patch token information after feature merging. Extensive experiments on ASVspoof 2021 LA, DF, and In-the-Wild datasets demonstrate state-of-the-art performance with equal error rates of 0.88\%, 1.85\%, and 6.30\% respectively, substantially outperforming existing methods. Comprehensive ablation studies validate that the dual-branch architecture provides complementary discriminative information, Squeeze-and-Excitation Aggregation significantly improves SSL feature integration, and the combination of DWConv and SE modules is critical for effective class token enhancement. The superior performance on real-world scenarios demonstrates strong generalization capability to diverse acoustic conditions and unseen spoofing attacks. The source code is available at \href{https://github.com/tuandattt/XLSR-Ebranchformer.git}{this link}.
\end{abstract}

\section{Introduction}
\label{sec:intro}

In recent years, sophisticated voice synthesis techniques such as voice conversion and text-to-speech (TTS) have advanced rapidly, posing significant threats to the integrity of voice-based authentication systems. As a result, guaranteeing the trustworthiness and robustness of speech-based systems has emerged as a critical imperative. Leveraging the rich representations from speech foundation models such as wav2vec 2.0 \cite{wav2vec2}, HuBERT \cite{hubert}, and WavLM \cite{wavlm}-models that have driven notable advances in automatic speech recognition (ASR) \cite{asr-example}, speaker verification (SV) \cite{espnet}, and other speech applications-could significantly improve the robustness and generalization of anti-spoofing systems.

However, a critical challenge lies in designing architectures capable of effectively processing and exploiting the representation features from self-supervised learning (SSL) models, which offer high-dimensional feature outputs. A series of convolution-based and Transformer-based architectures have been employed as downstream models \cite{xlsr-conformer, xlsr-tcm}, achieving considerable success in deepfake detection tasks. Among these, the Conformer \cite{conformer}, which combines convolution and self-attention sequentially, exhibits superior performance to the Transformer by processing local and global context together.

Concurrently, the emergence of novel Transformer-based architectures such as Branchformer \cite{branchformer} and E-Branchformer \cite{ebranchformer}, which perform the combination of convolution and self-attention in parallel branches, addresses certain limitations of the Conformer \cite{compare-branchformer}. These architectures have demonstrated superior performance over the Conformer in ASR and spoken language understanding (SLU) tasks \cite{compare-branchformer}, establishing their effectiveness in modeling both local and global dependencies. However, their potential for audio deepfake detection remains unexplored. In Branchformer and E-Branchformer, one branch employs self-attention to capture long-range dependencies, while the other utilizes an advanced multi-layer perceptron (MLP) to model local patterns. This parallel design offers greater flexibility, interpretability, and customizability compared to sequential architectures.

In this work, we propose the first application of E-Branchformer architecture to audio deepfake detection, fundamentally departing from existing approaches by explicitly disentangling global and local feature learning. Unlike the baseline XLSR-Conformer \cite{xlsr-conformer} which employs a single classification token, our novel architecture introduces dual class tokens: one dedicated to capturing global contextual information through the self-attention branch, and another specialized for local temporal patterns via the convolution branch. This disentangled design enables each class token to learn discriminative features specific to its respective domain. We further introduce a fusion mechanism combining depthwise convolution and Squeeze-and-Excitation modules to effectively integrate these complementary representations for robust synthetic audio detection. Our specific contributions include:

\begin{enumerate}
    \item Achieving state-of-the-art performance on the ASVspoof 2021 LA, 2021 DF, and In-the-Wild (ITW) datasets with equal error rates (EER) of 0.88\%, 1.85\%, and 6.30\%, respectively.
    \item An extensive investigation of various strategies for merging local and global features for deepfake detection tasks.
    \item A comprehensive analysis of the influence of global and local features in deepfake detection through ablation studies.
\end{enumerate}

\section{Preliminaries}

\subsection{Conformer}

The Conformer architecture \cite{conformer} integrates four key components: two position-wise feed-forward networks (FFNs), a multi-head self-attention (MHSA) mechanism, and a convolution module arranged in a sequential manner. Each component incorporates residual connections \cite{residual} and layer normalization \cite{layer-norm} applied in a pre-normalization configuration. The FFN modules utilize two linear transformations with a Swish activation function \cite{activateion-function} inserted between them. Following the Macaron-Net design \cite{macron-design}, two distinct FFNs with half-step residual connections are implemented. The MHSA mechanism differs from the original Transformer \cite{transformer} by adopting relative positional encodings. A distinguishing feature of the Conformer is its convolution module, which comprises a sequence of operations: an initial pointwise convolution, a gated linear unit (GLU) activation \cite{GELU}, a 1-D depthwise convolution layer, batch normalization \cite{batch-norm}, a Swish activation, and a final pointwise convolution layer.

\subsection{E-Branchformer}

The E-Branchformer architecture \cite{ebranchformer} represents an advanced iteration of the original Branchformer model \cite{branchformer}. It maintains the dual Macaron-style FFN configuration similar to Conformer. However, E-Branchformer distinguishes itself by employing two parallel processing branches positioned between the FFNs, following the design principle introduced in Branchformer. The first branch leverages MHSA to model long-range dependencies and global contextual information, while the second branch employs a convolutional gating multi-layer perceptron (cgMLP) \cite{cgmlp} to capture short-range dependencies and local patterns. The outputs from both branches are integrated through a merge operation consisting of concatenation, a 1-D depthwise convolution, and a linear projection layer. This merging strategy demonstrates superior effectiveness compared to the straightforward concatenation-projection approach utilized in the original Branchformer architecture.

\section{Proposed Architecture}

% \begin{figure*}[t]
%     \centering
%     \includegraphics[width=\linewidth]{ICME26_Ebranchformer/final.drawio.png}
%     \caption{The architecture of a) Conformer; b) E-Branchformer; c) Proposed Model}
%     \label{fig:architecture}
% \end{figure*}

\begin{figure*}[t]
    \centering
    \includegraphics[width=0.85\linewidth]{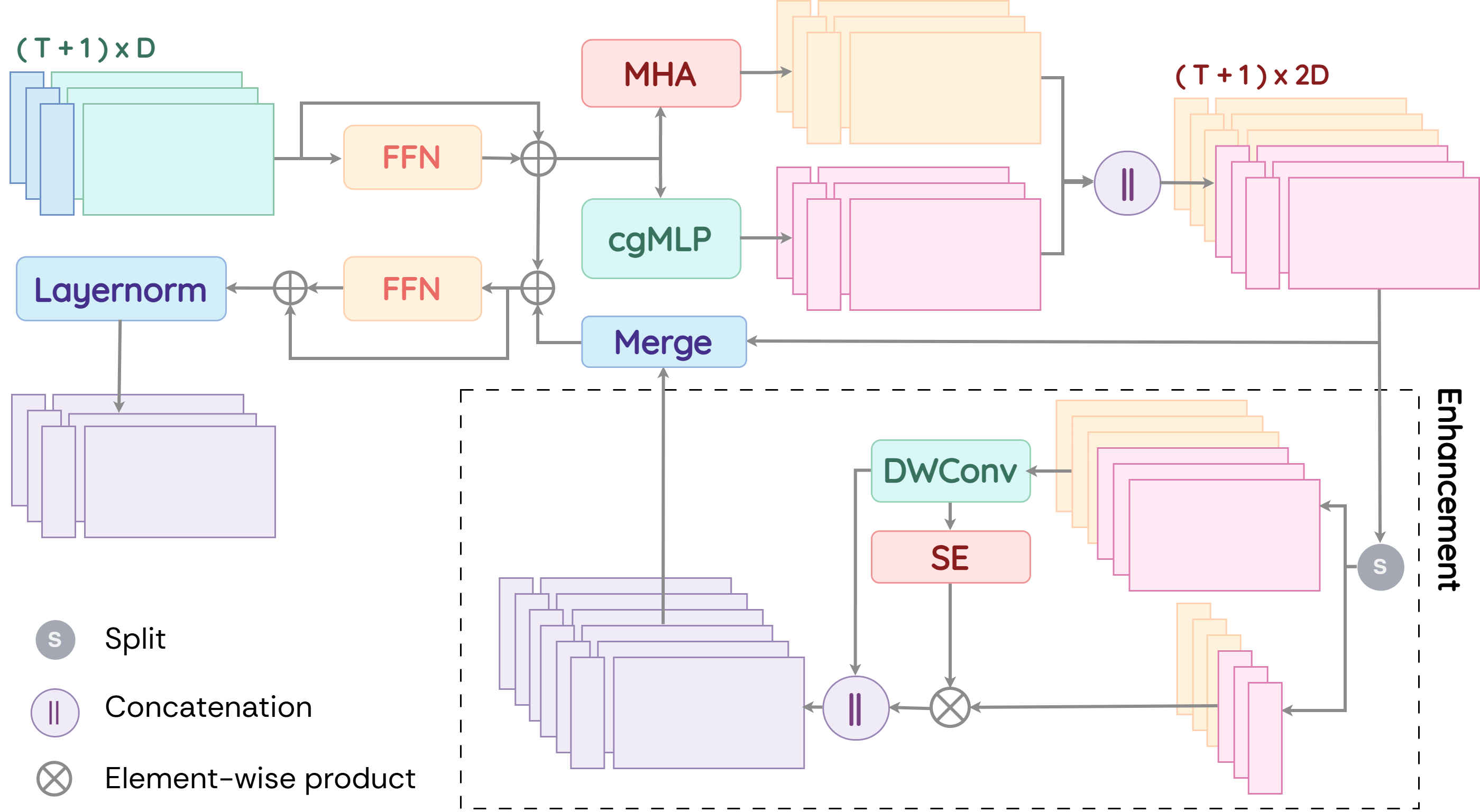}
    \caption{The architecture of E-Branchformer block used in the proposed encoder}
    \label{fig:architecture}
\end{figure*}

\subsection{Baseline Model}

We employ the state-of-the-art XLSR-Conformer \cite{xlsr-conformer} architecture as our baseline model. This baseline consists of two primary components: (i) a pre-trained XLS-R model \cite{xlsr}, derived from the wav2vec 2.0 framework \cite{wav2vec2}, serving as a feature extractor to derive contextualized representations from high-dimensional speech inputs; and (ii) a Conformer Encoder module. For a given input speech signal $O$, the SSL model generates output features of length $T$, denoted as $X = \text{SSL}(O) = (x_t \in \mathbb{R}^{D} \mid t = 1, \dots, T)$, where $D$ represents the dimensionality of the SSL model's output.

These extracted features $X$ undergo dimensionality reduction through an MLP layer equipped with a SeLU activation function prior to entering the Conformer Encoder. The resulting projected features are expressed as $\tilde{X} = \text{SeLU}(\text{Linear}(X))$, where $\tilde{X} = (\tilde{x}_t \in \mathbb{R}^{D'} \mid t = 1, \dots, T)$. The Conformer Encoder comprises $L$ stacked Conformer blocks, where each block integrates an MHSA mechanism and a Convolutional Module positioned between two feed-forward modules. To transform the sequence-to-sequence Conformer architecture into a classification framework, a learnable classification token is prepended to the encoder's input embeddings. The encoder input is denoted as $\tilde{X}_{\text{in}} = [\tilde{X}_{\text{CLS}}, \tilde{X}]$, where $\tilde{X} \in \mathbb{R}^{T \times D'}$, $\tilde{X}_{\text{CLS}} \in \mathbb{R}^{1 \times D'}$, and $[\cdot, \cdot]$ represents concatenation. Subsequently, the classification token's final state $\tilde{X}_{\text{CLS}}$ from the last Conformer block is passed through a linear classification layer to determine whether the input represents bonafide or spoofed speech. Throughout the training process, the classification token $\tilde{X}_{\text{CLS}}$ learns to encode discriminative features essential for distinguishing synthetic speech from authentic samples.

\subsection{Overall Proposed Model}
\label{sec:proposed}

Figure \ref{fig:architecture} illustrates the proposed architecture for audio deepfake detection. In alignment with recent advances in anti-spoofing systems, we adopt speech foundation models, specifically XLS-R, as our feature extraction front-end. These SSL-based models have consistently outperformed conventional acoustic feature extractors by delivering rich, contextualized representations of speech signals \cite{recursive-feature, voice-to-beat}.

Our architecture is composed of several interconnected components. Initially, the pre-trained XLS-R model processes the input speech waveform and produces multi-layer representations. The output speech representation exhibits a shape of $T \times D$, where $T$ represents the temporal sequence length and $D$ denotes the channel dimensionality of the XLS-R representation. To effectively harness information distributed across different transformer layers within the SSL model, we integrate an aggregation module that synthesizes features from multiple SSL layers. This aggregation mechanism can be realized through various approaches, ranging from simple weighted summation to more sophisticated attention-based techniques such as Squeeze-and-Excitation Aggregation (SEA) \cite{sea} and Attentive Merging (AttM) \cite{attentive-attention}.

Following aggregation, the XLS-R representation undergoes dimensional projection to $D'$ dimensions and is concatenated with a learnable classification token $\tilde{X}_{\text{CLS}}$ to construct the input sequence $\tilde{X}_{\text{in}} \in \mathbb{R}^{(T+1) \times D'}$ for the E-Branchformer model. This class token serves as a global representation accumulator specifically designed to facilitate the classification objective.

The core of our model consists of $L$ stacked E-Branchformer blocks, each featuring two parallel processing branches. The first branch incorporates a multi-head self-attention mechanism to model global dependencies and capture long-range contextual patterns across the temporal dimension of the speech signal. In parallel, the second branch employs a convolution-based architecture to extract local features and encode short-term temporal dependencies. This bifurcated design philosophy allows the model to concurrently process both global and local acoustic characteristics, which prove essential for discriminating synthetic speech from authentic samples.

Upon completion of the parallel processing through all $L$ blocks, the outputs from both branches undergo a fusion operation. The fused representation is subsequently decomposed into the class token and patch tokens. To further augment the discriminative capacity of the class token, the patch tokens traverse through a depth-wise convolution (DWConv) \cite{DWConv} layer followed by a Squeeze-and-Excitation (SE) \cite{SENetworks} module. The SE module computes channel-wise attention weights that accentuate the most salient features for the classification decision. These derived attention weights are then applied to the class token via element-wise multiplication, thereby effectively integrating the refined contextual information extracted from the patch tokens. Ultimately, the attention-weighted class token is merged with the recalibrated patch tokens, and the enriched class token serves as the input to a linear classification head that determines whether the input speech sample is bonafide or spoofed.

This architectural framework leverages the complementary strengths of attention-based global modeling and convolution-based local feature extraction, while preserving the flexibility to independently tune each branch for capturing distinct aspects of the audio signal pertinent to deepfake detection.

\section{Experiments}

\subsection{Datasets and Evaluation Metrics}

\begin{table}[t]
\centering
\caption{Datasets and number of samples used in our experiments.}
\resizebox{\columnwidth}{!}{%
\begin{tabular}{l c c c}
\hline
\textbf{Dataset} & \multicolumn{3}{c}{\textbf{Number of Samples}} \\
\cline{2-4}
& \textbf{Train} & \textbf{Valid} & \textbf{Test} \\
\hline
ASVspoof 2019 LA \cite{asv19} & 25,380 & 24,844 & - \\
ASVspoof 2021 LA \cite{asv21} & - & - & 181,566 \\
ASVspoof 2021 DF \cite{asv21} & - & - & 611,829 \\
ASVspoof 5 \cite{asvspoof5} & 182,357 & 140,950 & 680,774 \\
In-the-Wild (ITW) \cite{inthewild} & - & - & 31,779 \\
\hline
\end{tabular}%
}
\label{tab:datasets}
\end{table}

Our experimental evaluation encompasses two distinct dataset categories: (1) the ASVspoof series, which contains ASVspoof 2019 (LA19) \cite{asv19}, ASVspoof 2021 Logical Access (LA21), ASVspoof 2021 Deepfake (DF21) \cite{asv21}, and ASVspoof 5 \cite{asvspoof5}; and (2) the In-The-Wild (ITW) dataset \cite{inthewild}. Model performance is reported using the Equal Error Rate (EER), the primary metric of the ASVspoof Challenge series.

Table \ref{tab:datasets} summarizes the datasets and the number of samples utilized in our experiments. For the ASVspoof 2021 benchmarks, we employ LA19 as our training set, comprising 25,380 training samples and 24,844 validation samples. The evaluation is conducted on two test sets: LA21 containing 181,566 samples and DF21 with 611,829 samples, where the latter introduces additional complexity through lossy codec post-processing. ASVspoof 5 is the most large-scale benchmark in the series, comprising 182,357 training, 140,950 development, and 680,774 evaluation utterances, and models evaluated on it are trained on its own official training set. These datasets represent controlled laboratory conditions with diverse text-to-speech and voice conversion systems. Additionally, we evaluate our model on the ITW dataset, which consists of 31,779 test samples collected from real-world scenarios, providing a more challenging evaluation setting with various acoustic conditions and recording devices.

\subsection{Experimental Setup}

\begin{table}[t]
\centering
\caption{Hyperparameter configuration of the proposed model.}
\resizebox{\columnwidth}{!}{%
\begin{tabular}{lll}
\hline
\textbf{Group} & \textbf{Hyperparameter} & \textbf{Value} \\ \hline
\multirow{3}{*}{Front-end}
 & SSL model & XLS-R 300M \cite{xlsr} \\
 & Output dimensionality $D$ & 1024 \\
 & Fine-tuned during training & Yes \\ \hline
\multirow{4}{*}{Encoder}
 & E-Branchformer blocks $L$ & 4 \\
 & Embedding dimensionality $D'$ & 144 \\
 & Attention heads & 4 \\
 & Depthwise kernel size $\kappa$ & 31 \\ \hline
\multirow{6}{*}{Optimization}
 & Optimizer & Adam \cite{adam} \\
 & Learning rate & $1 \times 10^{-6}$ \\
 & Weight decay & $1 \times 10^{-4}$ \\
 & Batch size & 20 \\
 & Training epochs & 7 \\
 & Loss function & Weighted cross-entropy \\ \hline
\end{tabular}%
}
\label{tab:hyperparams}
\end{table}

% TODO: supply the remaining values once confirmed from the training logs -
% dropout rate, SE reduction ratio r, readout DWConv kernel size, random seed.

For models trained on the LA19 dataset, we adopt the training protocol established in the baseline system \cite{xlsr-conformer}\footnote{\url{https://github.com/ErosRos/conformer-based-classifier-for-anti-spoofing.git}}. Audio samples are preprocessed by either truncating or concatenating to generate fixed-length segments of approximately 4 seconds (equivalent to 64,600 samples at 16 kHz sampling rate) for both training and inference phases. Table \ref{tab:hyperparams} summarizes the complete hyperparameter configuration of the proposed model; the settings below are identical across all four benchmarks unless stated otherwise.

The XLS-R 300M model \cite{xlsr} serves as the front-end and is \emph{fine-tuned jointly} with the downstream network rather than kept frozen, so that its representations adapt to the spoofing-detection objective; its 1024-dimensional output is projected to $D' = 144$ before entering the encoder. The E-Branchformer encoder comprises $L = 4$ blocks, each employing 4 attention heads in the global branch, giving a head dimensionality of 36. The depthwise convolutions of the local branch use a kernel size of $\kappa = 31$, matching the receptive field of the Conformer convolution module in the baseline so that the two architectures are compared under equivalent local context.

We employ the Adam optimizer \cite{adam} with a learning rate of $1 \times 10^{-6}$, a weight decay coefficient of $1 \times 10^{-4}$, and a batch size of 20. Because bonafide utterances are substantially outnumbered by spoofed ones in the ASVspoof training partitions, the network is optimized with a weighted cross-entropy loss whose class weights are set inversely proportional to the class frequencies. To enhance model robustness and generalization, we incorporate RawBoost \cite{rawboost} data augmentation techniques following the methodology in \cite{xlsr-conformer}. Specifically, we apply the ``Algo5'' augmentation variant for the ASVspoof 2021 LA, ASVspoof 5, and ITW evaluations, and the ``Algo3'' variant for the ASVspoof 2021 DF evaluation. These augmentation schemes introduce realistic perturbations to the training data, simulating various acoustic conditions and transmission effects.

Each model is trained for a fixed budget of 7 epochs, and the checkpoint exhibiting the lowest validation loss is selected for final evaluation. For assessment on the ITW dataset, we use the same checkpoint as for the LA21 test set, ensuring consistency in cross-dataset evaluation.

\subsection{Overview of State-of-the-Art Models}

We compare our proposed approach against several recent state-of-the-art systems that leverage self-supervised learning features for audio deepfake detection. These methods predominantly utilize XLS-R or WavLM as feature extractors, coupled with various downstream architectures for discriminative feature learning. XLSR-Conformer+TCM \cite{xlsr-tcm} employs a Conformer encoder with temporal-channel modeling to capture both temporal and channel dependencies. XLSR-AASIST \cite{xlsr-aasist} integrates the AASIST architecture with XLS-R representations for enhanced spoofing detection. WavLM-MFA \cite{wavlm-mfa} utilizes WavLM features with a multi-fusion attentive classifier. XLSR-SLS \cite{sls} adopts a sensitive-layer-selection (SLS) classifier that adaptively weights the most informative SSL layers, while XLSR-MoE \cite{moe} employs a mixture-of-experts framework for ensemble learning. Among these methods, XLSR-Mamba \cite{xlsr-mamba} leverages bidirectional state space models with XLS-R features, representing one of the most competitive approaches in recent audio anti-spoofing research. These diverse architectures provide comprehensive benchmarks for evaluating the effectiveness of our proposed E-Branchformer-based detection system.

\subsection{Experimental Results}

\begin{table}[t]
\centering
\caption{Comparative performance analysis against state-of-the-art methods on the ASVspoof 2021 evaluation set and ITW dataset ($\dagger$ indicates our reproduction, \textbf{bold} highlights top performance, \underline{underline} marks runner-up results, and `-' signifies unavailable data)}
\resizebox{\columnwidth}{!}{%
\begin{tabular}{lccc}
\hline
\multirow{2}{*}{\textbf{Model}} & \multicolumn{3}{c}{\textbf{EER (\%)}}           \\ \cline{2-4}
                                & \textbf{21LA} & \textbf{21DF} & \textbf{ITW}  \\ \hline
XLSR-Conformer \cite{xlsr-conformer}$\dagger$            & 1.38           & 2.27           & 8.29          \\
XLSR-Conformer+TCM \cite{xlsr-tcm}$\dagger$              & 1.18           & 2.25           & 7.79          \\
XLSR-AASIST \cite{xlsr-aasist}                    & 1.00           & 3.69           & 10.46         \\
WavLM-MFA \cite{wavlm-mfa}                      & 5.08           & 2.56           & -             \\
XLSR-SLS \cite{sls}                       & 5.08           &  1.92           & 7.46          \\
XLSR-MoE \cite{moe}                       & 2.96           & 2.54           & 9.17          \\
XLSR-Mamba \cite{xlsr-mamba}                     & \underline{0.93}     & 1.88  &  \underline{6.71}    \\
Nes2Net \cite{nes2net}                           & 1.72           & 2.01           & 7.18          \\
Nes2Net-X \cite{nes2net}                         & 1.82           & \underline{1.87}  & 6.74       \\ \hline
Proposed model  & \textbf{0.88}     & \textbf{1.85}  &  \textbf{6.30}    \\ \hline
\end{tabular}%
}
\label{tab:baseline-results}
\end{table}

Table \ref{tab:baseline-results} presents the performance comparison between our proposed model and recent SOTA systems on the ASVspoof 2021 evaluation sets and the ITW dataset. Our proposed E-Branchformer-based model achieves SOTA performance across all three evaluation benchmarks, demonstrating substantial improvements over existing methods.

Specifically, our model attains an EER of 0.88\% on ASVspoof 2021 LA, 1.85\% on ASVspoof 2021 DF, and 6.30\% on the ITW dataset, establishing new performance benchmarks on all three datasets. No single competing system is the runner-up on all three sets: XLSR-Mamba previously achieved the best baseline result on 21LA (0.93\%) and ITW (6.71\%), whereas Nes2Net-X holds it on 21DF (1.87\%). Our approach surpasses whichever system is strongest on each set, delivering relative improvements of 5.4\% over XLSR-Mamba on 21LA and 1.1\% over Nes2Net-X on 21DF. More notably, on the challenging out-of-domain ITW dataset, our model outperforms XLSR-Mamba by achieving 6.30\% EER compared to 6.71\% EER, a relative improvement of 6.1\%, and does so while also improving on the second-best ITW system, Nes2Net-X, at 6.74\%. This substantial gain on the real-world ITW dataset demonstrates the superior generalization capability of our architecture to unseen spoofing attacks and diverse acoustic conditions.
When compared against the XLSR-Conformer+TCM baseline, which we reproduced and serves as our direct baseline, our proposed model achieves remarkable relative EER reductions of 25.4\% on 21LA (from 1.18\% to 0.88\%), 17.8\% on 21DF (from 2.25\% to 1.85\%), and 19.1\% on ITW (from 7.79\% to 6.30\%). These consistent improvements across all evaluation sets validate the effectiveness of our dual-branch architecture in capturing both global and local discriminative features for deepfake detection.

The comparison that speaks most directly to the choice of backbone is the one against XLSR-Conformer, which shares our XLS-R front-end, our training protocol and our data augmentation, and differs principally in how local and global modelling are arranged. Replacing the sequential Conformer encoder with our E-Branchformer-based design lowers the EER from 1.38\% to 0.88\% on 21LA (36.2\% relative), from 2.27\% to 1.85\% on 21DF (18.5\%), from 8.29\% to 6.30\% on ITW (24.0\%), and, in Table \ref{tab:asv5_results}, from 6.19\% to 5.44\% on ASVspoof 5 (12.1\%). The advantage therefore holds on all four benchmarks, spanning matched (21LA), codec-degraded (21DF), out-of-domain real-world (ITW) and large-scale (ASVspoof 5) conditions, which makes a benchmark-specific artifact an implausible explanation. It is moreover widest precisely where the evaluation is furthest from the training distribution-1.99 percentage points on ITW against 0.50 on 21LA-indicating that parallel global-local modelling produces representations that transfer better rather than merely fitting the in-domain condition more tightly. The same ordering persists against XLSR-Conformer+TCM, an enhanced Conformer whose temporal-channel modelling module was introduced specifically to strengthen the class token: our model remains ahead of it on every benchmark, which indicates that the gap cannot be closed simply by improving the readout of a sequential backbone.

We stress that this contrast is between two complete systems, and therefore reflects the dual class tokens and the DWConv--SE fusion described in Section \ref{sec:proposed} in addition to the encoder itself; the contribution of each component is isolated in Section \ref{sec:ablation}. The property attributable to the backbone alone is examined directly in Section \ref{sec:diagonality}, where the attention behaviour of the two encoders is measured under identical inputs.

Furthermore, our model substantially outperforms other competitive systems including XLSR-AASIST and WavLM-MFA, achieving relative improvements of up to 12\% and 82.7\% on the 21LA dataset respectively. The consistent superior performance across controlled laboratory conditions (21LA, 21DF) and real-world scenarios (ITW) underscores the robustness and practical applicability of our approach. These results demonstrate that the parallel processing of global and local features through the E-Branchformer architecture, combined with the SE-based feature recalibration mechanism, provides a more effective framework for audio deepfake detection compared to sequential processing architectures like Conformer or other existing approaches.

\begin{table}[t]
\centering
\caption{Comparison with SOTA single systems on the ASVspoof 5 dataset ($\dagger$ indicates our reproduction; the remaining baselines are quoted from \citet{nes2net}).}
\label{tab:asv5_results}
\resizebox{\columnwidth}{!}{%
\begin{tabular}{lc}
\hline
\textbf{System} & \textbf{EER (\%)} \\
\hline
XLSR-AASIST \cite{xlsr-aasist} & 6.08 \\
XLSR-Conformer \cite{xlsr-conformer}$\dagger$ & 6.19 \\
XLSR-Conformer+TCM \cite{xlsr-tcm}$\dagger$ & 6.03 \\
XLSR-Mamba \cite{xlsr-mamba}$\dagger$ & 6.24 \\
Nes2Net \cite{nes2net} & 6.13 \\
Nes2Net-X \cite{nes2net} & \underline{5.92} \\
\hline
Proposed model & \textbf{5.44} \\
\hline
\end{tabular}%
}
\end{table}

Table \ref{tab:asv5_results} compares our model against state-of-the-art single systems on the ASVspoof 5 benchmark. Our proposed model achieves an EER of 5.44\%, surpassing all competing systems. In particular, it outperforms the strongest baseline, Nes2Net-X, by 0.48 percentage points (5.92\% to 5.44\%, an 8.1\% relative reduction), improves on our reproduced XLSR-Mamba by 0.80 percentage points (6.24\% to 5.44\%), and reduces the EER of the reproduced XLSR-Conformer+TCM from 6.03\% to 5.44\%, confirming that the dual CLS token design generalizes effectively beyond the ASVspoof 2021 evaluation conditions.

\section{Ablation Study}
\label{sec:ablation}

\subsection{Impact of Global and Local Branches}

\begin{figure}[t]
\centering
\begin{tikzpicture}
\begin{axis}[
    width=\columnwidth,
    height=5.6cm,
    ybar=1.4pt,
    bar width=7.5pt,
    enlarge x limits=0.16,
    ymin=0, ymax=11,
    ytick={0,2,4,6,8,10},
    ylabel={EER (\%)},
    ylabel near ticks,
    symbolic x coords={21LA,21DF,ITW,ASVspoof 5},
    xtick=data,
    ymajorgrids=true,
    grid style={seriesgrid,line width=0.4pt},
    axis x line*=bottom,
    axis y line*=left,
    axis line style={seriesaxis},
    tick style={seriesaxis},
    tick align=outside,
    xtick style={draw=none},
    label style={font=\footnotesize},
    tick label style={font=\footnotesize},
    legend style={
        at={(0.5,1.02)}, anchor=south, legend columns=3,
        draw=none, fill=none, font=\scriptsize, column sep=5pt,
        /tikz/every even column/.append style={column sep=3pt}},
    legend image code/.code={\draw[#1] (0cm,-0.06cm) rectangle (0.22cm,0.11cm);},
    nodes near coords,
    every node near coord/.append style={
        font=\tiny, color=black!70, rotate=90, anchor=west, inner sep=1.6pt,
        /pgf/number format/.cd, fixed, fixed zerofill, precision=2},
]
\addplot[fill=seriesone,draw=none] coordinates
    {(21LA,1.92) (21DF,2.54) (ITW,8.70) (ASVspoof 5,7.02)};
\addplot[fill=seriestwo,draw=none] coordinates
    {(21LA,0.94) (21DF,2.29) (ITW,6.73) (ASVspoof 5,6.16)};
\addplot[fill=seriesthree,draw=none] coordinates
    {(21LA,0.88) (21DF,1.85) (ITW,6.30) (ASVspoof 5,5.44)};
\legend{Local only, Global only, Proposed}
\end{axis}
\end{tikzpicture}
\caption{Impact of the global and local feature extraction branches on detection performance. Bars report EER (\%), for which lower is better, for the local branch alone, the global branch alone, and the proposed dual-branch model on the four evaluation benchmarks.}
\label{fig:ablation-branches}
\end{figure}
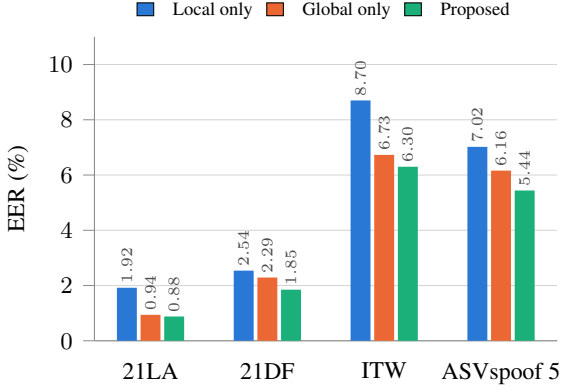

Figure \ref{fig:ablation-branches} examines the individual contribution of each branch. The global branch alone consistently outperforms the local branch across all benchmarks, suggesting that long-range contextual dependencies are more discriminative for spoofing detection than local temporal patterns. Nevertheless, the local branch provides complementary cues that, when combined via dual CLS tokens, yield substantial gains over either branch in isolation-reducing EER from 0.94\% to 0.88\% on 21LA, 2.29\% to 1.85\% on 21DF, 6.73\% to 6.30\% on ITW, and 6.16\% to 5.44\% on ASVspoof 5-demonstrating that both global and local features are essential for robust deepfake speech detection.

\subsection{Impact of SSL Feature Aggregation Strategies}

\begin{table}[t]
\centering
\caption{Impact of SSL feature aggregation strategies on detection performance.}
\resizebox{\columnwidth}{!}{%
\begin{tabular}{lcccc}
\hline
\multirow{2}{*}{\textbf{Model}} & \multicolumn{4}{c}{\textbf{EER (\%)}}           \\ \cline{2-5}
& \textbf{21LA} & \textbf{21DF} & \textbf{ITW} & \textbf{ASVspoof 5} \\ \hline
AttM \cite{attentive-attention}  & 0.98  & 2.29 & 8.76 & 6.32 \\
Weighted sum  & 1.07 & 2.33  & 8.82  & 6.18 \\
Last hidden layer & 1.00 & 2.46  & 8.67 & 6.23 \\
\hline
SEA \cite{sea} & \textbf{0.88}     & \textbf{1.85}  &  \textbf{6.30}  & \textbf{5.44}  \\ \hline
\end{tabular}%
}
\label{tab:ablation-aggregation}
\end{table}

Table \ref{tab:ablation-aggregation} compares different strategies for aggregating SSL layer representations. Our adopted SEA \cite{sea} achieves the best performance across all benchmarks with EERs of 0.88\%, 1.85\%, 6.30\%, and 5.44\% on 21LA, 21DF, ITW, and ASVspoof 5 respectively, outperforming AttM \cite{attentive-attention}, weighted sum, and last hidden layer aggregation. The consistent margin over simpler alternatives such as weighted sum and last hidden layer indicates that an adaptive, content-aware selection of SSL layers is critical for extracting the most informative representations for spoofing detection.

\subsection{Impact of Key Architectural Components}

\begin{table}[t]
\centering
\caption{Ablation study on DWConv and SE module contributions.}
\resizebox{\columnwidth}{!}{%
\begin{tabular}{lcccc}
\hline
\multirow{2}{*}{\textbf{Model}} & \multicolumn{4}{c}{\textbf{EER (\%)}}           \\ \cline{2-5}
& \textbf{21LA} & \textbf{21DF} & \textbf{ITW} & \textbf{ASVspoof 5} \\ \hline
Proposed model & \textbf{0.88}     & \textbf{1.85}  &  \textbf{6.30}  & \textbf{5.44}  \\ \hline
w/o SENet & 0.89  & 2.19 & 9.34 & 5.57 \\
w/o DWConv  & 0.98 & 2.54  & 9.87 & 6.05 \\
w/o DWConv and SENet  & 1.06 & 2.87  & 10.90 & 7.18 \\ \hline
\end{tabular}%
}
\label{tab:ablation-components}
\end{table}

Table \ref{tab:ablation-components} ablates the two components of our fusion mechanism. Removing either SENet or DWConv consistently degrades performance across all benchmarks, and removing both yields the worst results with EERs of 1.06\%, 2.87\%, 10.90\%, and 7.18\% on 21LA, 21DF, ITW, and ASVspoof 5 respectively-representing absolute degradations of up to 4.60\% on ITW compared to the full model. DWConv contributes more significantly than SENet, particularly on 21DF and ITW, indicating its critical role in capturing local cross-token interactions. Together, both components are complementary and jointly essential for achieving the best performance of our proposed fusion mechanism.

\subsection{Layer-Wise Analysis of Global Branch}
\label{sec:diagonality}

\begin{figure*}[t]
    \centering
    \includegraphics[width=\linewidth]{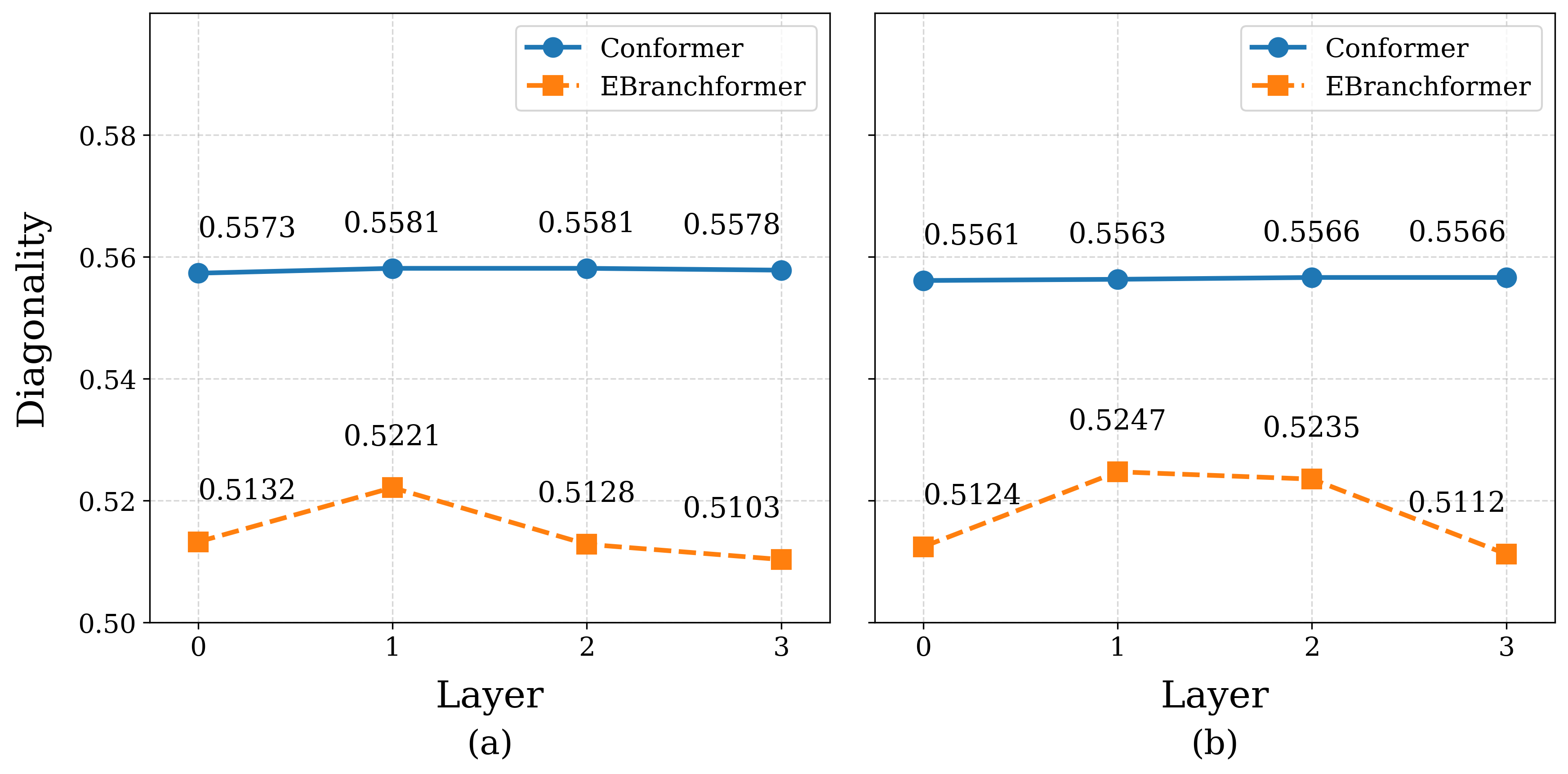}
    \caption{Diagonality of self-attention weights across encoder layers for Conformer and E-Branchformer. (a) ASVspoof 2021 LA. (b) ASVspoof 2021 DF.}
    \label{fig:diagonality}
\end{figure*}

To further validate the benefit of disentangling global and local processing, we analyze the self-attention patterns of both architectures using the diagonality score \cite{diagonality}, which quantifies the locality of attention: a higher score indicates that attention weights concentrate along the diagonal, reflecting a bias toward local context, while a lower score reflects a more globally distributed attention pattern.

As shown in Figure \ref{fig:diagonality}, the Conformer maintains consistently high diagonality scores across all layers ($\sim$0.556--0.558 on LA, $\sim$0.556 on DF), indicating that its self-attention is dominated by local context throughout the network. In contrast, E-Branchformer exhibits substantially lower diagonality scores ($\sim$0.510--0.522 on LA, $\sim$0.511--0.525 on DF), demonstrating that by offloading local feature extraction to a dedicated convolutional branch, the self-attention branch is freed to focus on broader, global dependencies. This architectural disentanglement allows E-Branchformer to capture global spoofing cues more effectively than the Conformer, which must simultaneously handle both local and global patterns within a single self-attention mechanism.

Two features of this measurement make it a statement about the architectures rather than about a particular dataset. First, the separation is present at \emph{every} encoder layer and on \emph{both} evaluation sets, with the two score ranges never overlapping; the effect is thus systematic rather than confined to a few layers where it might reflect optimisation noise. Second, it is obtained with the two encoders receiving identical XLS-R inputs and trained under an identical protocol, so the difference in attention behaviour is not inherited from the front-end. Read together with the EER comparison earlier in this section, the two observations support a single account: the Conformer expends part of its attention capacity on context its convolution module has already modelled, whereas delegating local extraction to a parallel branch leaves the E-Branchformer free to spend that capacity on the long-range inconsistencies-prosodic discontinuities and utterance-level spectral artifacts-that distinguish synthesised speech, which in turn is what the larger margins on the out-of-domain ITW and ASVspoof 5 conditions reflect.

\section{Conclusion}

In this work, we proposed a novel E-Branchformer-based architecture for audio deepfake detection that effectively leverages self-supervised speech representations through parallel processing of global and local features. Our model achieves SOTA performance with EERs of 0.88\%, 1.85\%, and 6.30\% on ASVspoof 2021 LA, DF, and In-the-Wild datasets respectively, substantially outperforming existing methods. Comprehensive ablation studies validate our key design choices: the dual-branch architecture with both attention-based and convolution-based processing provides complementary discriminative information, Squeeze-and-Excitation Aggregation (SEA) significantly outperforms alternative SSL feature integration strategies, and the combination of DWConv and SE modules is critical for effective class token enhancement with patch token information. The superior performance on the challenging In-the-Wild dataset demonstrates strong generalization capability to real-world scenarios with diverse acoustic conditions and unseen spoofing attacks. Future work will explore extending this architecture to multi-modal deepfake detection and investigating robustness against adversarial attacks.

\newpage

% Bibliography
\bibliography{icme2026references}

\end{document}